\documentclass{article}

\usepackage{enumitem}
\usepackage{microtype}
\usepackage{graphicx}
\usepackage{subfigure}
\usepackage{booktabs} 
\usepackage{xcolor}
\usepackage{xspace}
\usepackage{algorithm}
\usepackage{algpseudocode}
\usepackage{lineno}
\usepackage{hyperref}

\usepackage[accepted]{mlsys2025}

\usepackage{amsmath}   
\usepackage{amssymb}   
\usepackage{mathtools} 
\newcommand{\ourmethod}{GhostServe\xspace}
\newcommand{\minisection}[1]{\vspace{0.05in}\noindent {\bf #1}}

\mlsystitlerunning{\ourmethod: A Lightweight Checkpointing System in the Shadow for Fault-Tolerant LLM Serving}

\begin{document}

\twocolumn[


\mlsystitle{\ourmethod: A Lightweight Checkpointing System in the Shadow for Fault-Tolerant LLM Serving}


\mlsyssetsymbol{equal}{*}
\mlsyssetsymbol{lead}{$\dagger$}

\begin{mlsysauthorlist}
\mlsysauthor{Shakya Jayakody}{equal,ucf}
\mlsysauthor{Youpeng Zhao}{equal,lead,ucf}
\mlsysauthor{Chinmay Nehate}{ucf}
\mlsysauthor{Jun Wang}{ucf}

\end{mlsysauthorlist}

\mlsysaffiliation{ucf}{University of Central Florida, Orlando, FL, USA}

\mlsyscorrespondingauthor{Youpeng Zhao}{youpeng.zhao@ucf.edu}

\mlsyskeywords{Machine Learning, MLSys}

\vskip 0.3in

\begin{abstract}
The rise of million-token, agent-based applications has placed unprecedented demands on large language model (LLM) inference services. 
The long-running nature of these tasks increases their susceptibility to hardware and software faults, leading to costly job failures, wasted resources, and degraded user experience.
The stateful key-value (KV) cache, which grows with the sequence length, presents a central challenge as it is a critical and vulnerable component in distributed serving systems.
In this work, we propose \textbf{\ourmethod}, a novel checkpointing solution to facilitate fault-tolerant LLM serving.
Specifically, \ourmethod protects the streaming KV cache \textit{in the shadow} by applying erasure coding to generate and store the parity shards in host memory.
In the event of device failures, \ourmethod enables fast reconstruction of the lost KV cache, allowing the inference process to resume seamlessly without costly full recomputation or state replication.
Evaluations demonstrate that \ourmethod reduces checkpointing latency by up to 2.7$\times$ and recovery latency by 2.1$\times$ for a single batch, and 1.2$\times$ median response latency compared to existing methods, in the presence of system failures, paving the way for high-availability and cost-effective LLM serving at scale.
\end{abstract}
]

\printAffiliationsAndNotice{\mlsysEqualContribution \, \mlsysProjectLead} 

\section{Introduction}
\label{sec:intro}
The development of large language models (LLMs) has brought the prospect
of artificial general intelligence (AGI) within closer reach, due to their exceptional performance across diverse language understanding and generation tasks~\cite{brown2020language,openai2025gptoss,llama,guo2025deepseek}.
A key driving force behind this rapid progress is the prevalence of scaling laws, which indicate that the performance of LLMs improves as the model size increases~\cite{scaling,scaling2}.
Naturally, the deployment of inference serving has become increasingly popular in distributed environments, where specialized accelerators such as GPUs or TPUs are utilized to accommodate these compute- and memory-intensive workloads~\cite{tpu,vLLM,zheng2024sglang,agrawal2023sarathi,Qin2025MooncakeTM}.
Subsequently, the power of LLMs has been continually pushed to the next level, where agent-based applications like GPT Codex~\cite{gptcodex}, and Claude Code~\cite{claudecode} are capable of performing million-token complex coding tasks, further streamlining and accelerating scientific and engineering development.

However, as the intensity of LLM serving workloads rapidly increases, hardware or software errors are often unavoidable, especially in large-scale distributed environments.
Interruptions are common and frequent in enterprise-level high-performance computing (HPC) data centers~\cite{mohan2021checkfreq,gemini,bytecheckpoint,burstgpt,moetion}.
With the emergence of LLM-based AI agent applications, a single point of failure could result in catastrophic consequences for the entire workflow, degraded user experiences, and even huge financial losses if not promptly recovered~\cite{chatgpt1,crowdstrike1,awsoutage}.
Recent studies highlight the importance of resilient LLM services and the dire need for inference engine and infrastructure improvements to reduce mitigation time and impact~\cite{reliabileai,demistifying,burstgpt}.
Furthermore, the autoregressive characteristic of LLM inference requires storing crucial but growing transient intermediate states, e.g., key-value (KV) cache for efficient token generation~\cite{alisa,fairseq,vLLM}. 
In the event of a failure, this volatile state is lost, forcing the system to restart the inference job from the very beginning. 
For long-context tasks, this implies a costly recomputation of the entire KV cache and can take as long as tens of minutes~\cite{zhu2025sampleattention,cp}.
Therefore, it has become increasingly essential to promote real-world system reliability and improve cost-effectiveness for LLM serving at scale~\cite{reliabileai,demistifying,medha}.

\renewcommand{\arraystretch}{1.02}
\begin{table*}[!t]
\begin{center}
\caption{Comparison of prior works and \ourmethod.}
\vspace{1mm}
\resizebox{.99\linewidth}{!}{
\begin{tabular}{l||c|c|c|c||c}
\toprule
 & Checkpointing & Recomputation & Backup Overhead & Recovery Overhead & Overall Costs \\ \midrule\midrule
 \textbf{
  SGLang~\cite{zheng2024sglang}
  } & $\times$ & $\surd$ & Low & High & High \\
  \textbf{
  SGLang + Replication~\cite{dejavu}
  } & $\surd$ & $\times$ & High & Medium & High \\
  \textbf{GhostServe (Ours)} & $\surd$ & $\surd$ & Low & Low & Low \\
\bottomrule
\end{tabular}
\label{tab:1}
}
\end{center}
\end{table*}

Although there have been prior research efforts to address the fault-tolerant issues for LLM applications~\cite{gemini,bytecheckpoint,pccheck,dejavu,fastertransformer,robustlt},  such designs often fall short in the case of LLM serving.
\underline{First}, previous systems are generally optimized for offline LLM training, where the workloads are often homogeneous and predictable.
The protection of model weights and optimizer states can be readily achieved via low-latency concurrent checkpointing~\cite{pccheck,bytecheckpoint,moetion}.
In the case of LLM serving, due to the variety of input prompts and output lengths, the computation procedure is more dynamic, thus harder to determine and profile beforehand.
\underline{Second}, applying naive checkpointing for LLM serving could induce severe latency overheads, due to the growing memory footprint of the KV cache.
While methods like D{\'e}j{\`a}Vu propose to alleviate such overhead by overlapping the I/O with computation in pipeline parallelism, it does not work well in intra-node tensor parallelism for real-time LLM serving~\cite{dejavu}.
Moreover, replication of the entire KV cache often leads to significant host memory overhead, as the KV cache size can be as big as hundreds of gigabytes (GBs) for million-token contexts.
The lack of sufficient node memory could lead to CPU oversubscription, further downgrading the system throughput performance and predictability~\cite{oversub1,oversub2}.

\begin{figure}[!t]
     \centering
         \includegraphics[width=\linewidth]{ 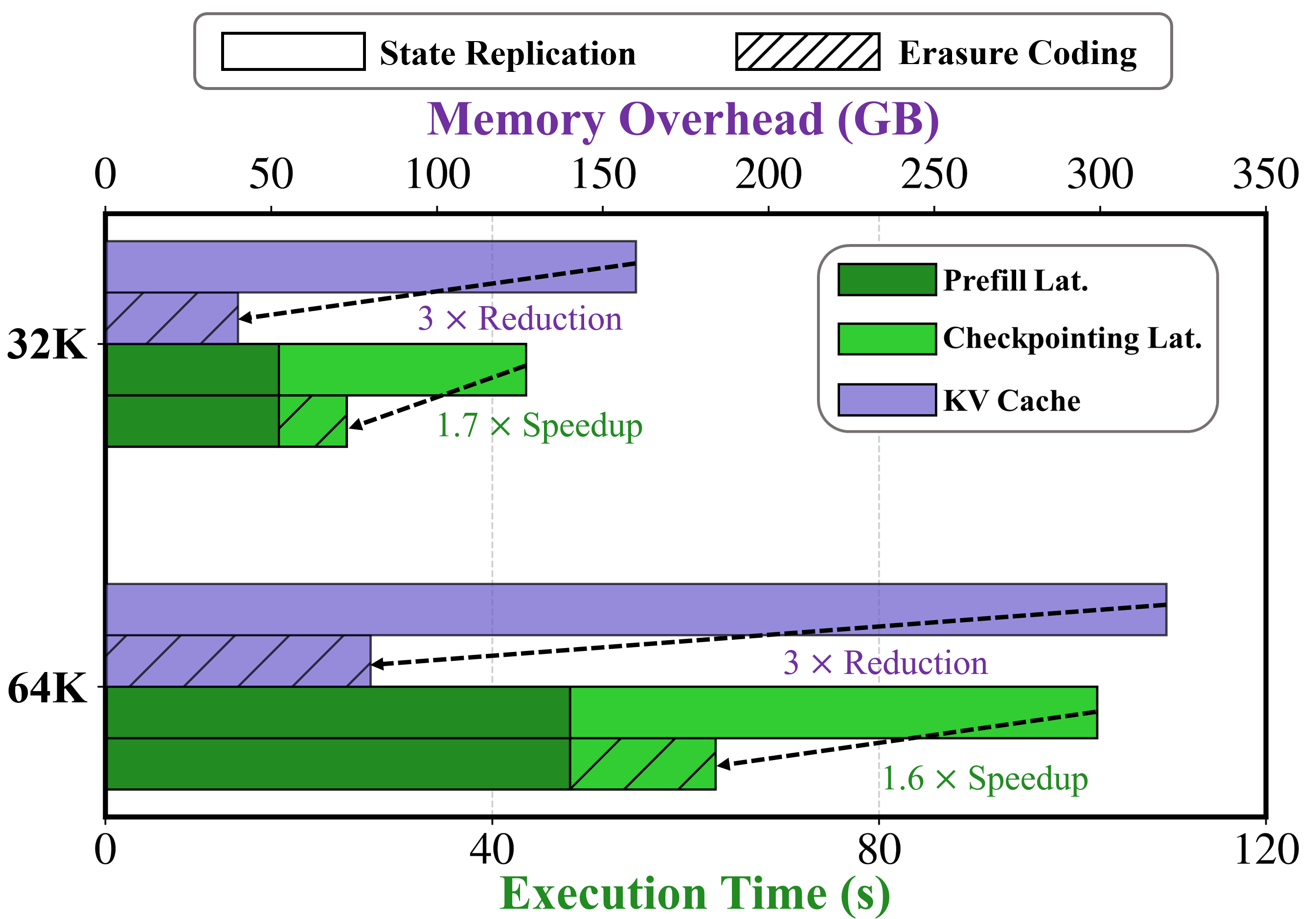}
         \vspace{-8mm}
        \caption{
        Comparison of checkpointing latency and memory overhead of our erasure coding (8:2) method and state replication~\cite{dejavu} during prefill.
        Results are profiled with LLaMA-3-70B using SGLang with a batch size of 16, and varying input sequence lengths (32/64K), a chunk size of 2K, in tensor parallelism (TP=8) across 8$\times$H200 GPUs.
} 
	\label{fig:motivation2}
\end{figure}

In this work, inspired by the idea of erasure coding~\cite{li2013erasure,aguilera2005using}, we propose \textbf{\ourmethod}, a new lightweight checkpointing framework to facilitate fault-tolerant LLM serving in the wild.
The concept of erasure coding has been widely implemented in modern distributed storage systems, where data is split and encoded to generate redundant data shards for data recovery in the event of complete disk failures~\cite{huang2012erasure,li2013erasure,kv1,kv2}.
\textbf{
The key insight behind \ourmethod is that instead of directly protecting the entire KV cache, we only need to protect the encoded redundant pieces in the host memory, thus significantly reducing the I/O transfer latency and memory overheads.
}
As shown in Figure~\ref{fig:motivation2}, our method reduces the host memory overhead and checkpointing latency by 75\% and 73\%, respectively.
However, achieving such improvements demands solving two unique challenges.

First, applying erasure coding is not straightforward.
Erasure codes are often defined over binary fields, making them incompatible with the floating-point (FP) representation of the KV cache.
To address this issue, we adopt an integer-centric view of the KV cache and develop a suite of highly optimized GPU kernels to perform lossless encoding operations with low latency, supporting a range of standard codes, like XOR~\cite{aguilera2005using}, RDP~\cite{corbett2004row}, and Reed-Solomon~\cite{rs}.
Thanks to such implementations, \ourmethod silently operates \textit{in the shadow}, induces minimal overhead in both checkpointing and recovery processes.

Second, during distributed serving, each worker independently generates its portion of the KV cache. 
A strawman solution of applying erasure coding to a distributed KV cache encounters the issue of GPU memory overhead, degrading LLM serving throughput.
To this end, \ourmethod conducts the erasure coding at the granularity of a chunk (group of tokens), and assigns a dedicated worker for parity generation.
Such a design removes the GPU memory overhead, while offering flexibility and fine-grained fault tolerance for users, and can be further combined with a recomputation strategy to foster faster recovery time. 
Moreover, to ensure system stability and steady GPU utilization, \ourmethod features a workload balancing strategy to rotate encoding assignment in a round-robin manner, where each GPU performs the encoding operation for each chunk at a time.
As shown in Table~\ref{tab:1}, compared against prior methods, \ourmethod provides a much more cost-effective checkpointing solution towards million-token LLM serving, with much lower system overheads.

\begin{figure*}[!t]
     \centering
         \includegraphics[width=\linewidth]{ 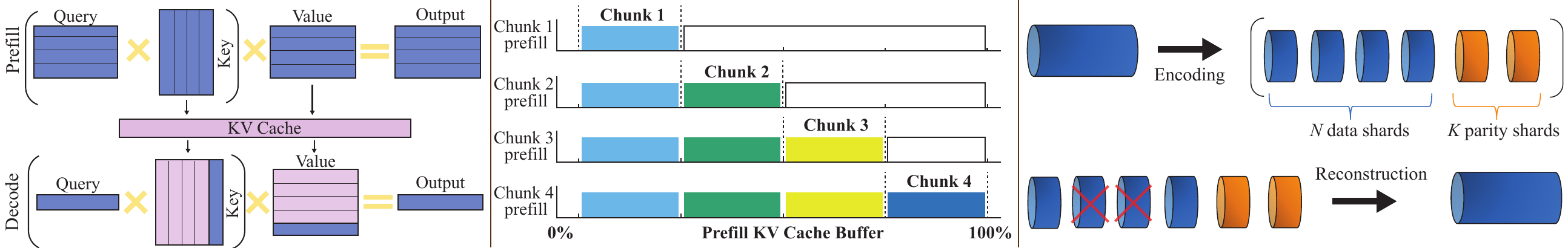}
        \vspace{-6mm}
        \caption{(a) Left: Prefill and decode stages in LLM inference. 
        In the prefill stage, all queries are processed with keys and values computed and stored in memory. 
        During the subsequent decode stage, only the attention for new queries is needed, significantly reducing attention computation. 
        (b) Middle: Chunked prefill mechanism. 
        During prefill, the input is divided into chunks of tokens, where the corresponding KV cache is written sequentially into contiguous chunks of the KV cache buffer. 
        (c) Right: Illustrative example of erasure coding.
        The input data is first partitioned into $N$ shards, which are then encoded to generate $K$ parity shards. 
        In the event of failures, the lost shards can be reconstructed from parity shards and the surviving data shards.
        }
     \label{fig:bg}
\end{figure*}

To summarize, this paper makes the following contributions:
\begin{itemize}
    \item We identify fault tolerance issues for LLM serving and propose a novel lightweight checkpointing system based on erasure coding, termed \ourmethod.
    \item We develop a suite of specialized GPU kernels to support KV cache encoding and reconstruction with minimal latency overhead.
    \item \ourmethod performs checkpointing at the chunk-level for distributed KV cache and a rotating workload balancing method to ensure minimal overhead and system stability.
    \item We implement \ourmethod in distributed settings on top of SGLang, and evaluate it across diverse workload scenarios.
    Experiments demonstrate that \ourmethod significantly improves the fault-tolerance capabilities of existing LLM serving systems, offering substantial checkpointing latency reduction and recovery time speedup.
\end{itemize}

\section{Background}
\minisection{LLM Inference.}
Inference in transformer-based large language models (LLMs)~\cite{vaswani2017attention} typically follows a two-phase pipeline: the prefill stage and the decode stage~\cite{efficientscaling}, as shown in Figure~\ref{fig:bg}~(a).
In the prefill stage, the model processes the entire input prompt in parallel, generating the key-value (KV) tensors, often known as the KV cache~\cite{fairseq}.
For prompts that often span thousands, and even millions of tokens, chunked-prefill is employed, where the input is partitioned into fixed-size chunks that are processed sequentially~\cite{zheng2024sglang, agrawal2023sarathi}.
This can reduce peak GPU memory usage, allowing for processing very long input prompts without running out of memory.
Furthermore, chunking creates opportunities for parallelism, where systems can begin the memory-bound decoding work at the same time, thus improving overall GPU utilization.
The mechanism of chunked-prefill is shown in Figure~\ref{fig:bg}~(b).
Once prefill completes, the decode stage begins, where the model generates tokens one-at-a-time, making it memory-bound instead of compute-bound.

\minisection{Distributed Serving.}
The immense memory footprint of modern LLMs, driven by both billion-level parameter weights and the dynamic KV cache for long contexts, has made it infeasible to serve them on a single accelerator. 
Consequently, distributed serving has become a necessity, leveraging model parallelism techniques~\cite{shoeybi2019megatron,huang2019gpipe}. 
A common practice is to employ tensor parallelism for intra-node scenarios, where each operator weight is split across multiple accelerators, with each device executing part of the computation in parallel with device-device synchronization at each transformer layer~\cite{shoeybi2019megatron}.
Specialized serving systems have introduced further optimization to address the overhead of the growing KV cache~\cite{vLLM,Qin2025MooncakeTM,zheng2024sglang}. 
vLLM proposes PagedAttention to virtually eliminate memory fragmentation in the KV cache, significantly improving throughput and memory efficiency~\cite{vLLM}.
MoonCake further develops a KVCache-centric architecture to leverage GPU, CPU, and SSD for efficient disaggregated KV cache at scale~\cite{Qin2025MooncakeTM}.
In terms of fault tolerance, the prevalent strategies for most serving systems are recomputation~\cite{fastertransformer}.
While D{\'e}j{\`a}Vu leverages replication method to perform concurrent KV cache `checkpointing'~\cite{dejavu}, it does not work well for existing 
high-performance serving systems, such as SGLang~\cite{zheng2024sglang}, especially in intra-node tensor parallelism.

\minisection{Erasure Coding.}
Erasure coding has long been a cornerstone of reliability in large-scale distributed storage systems, offering robust protection against hardware failures while significantly reducing the storage overhead of naive replication~\cite{huang2012erasure,li2013erasure}. 
The core idea is to generate parity shards using existing data, creating redundancy, as shown in Figure~\ref{fig:bg}~(c).
Among its variants, exclusive-OR (XOR) coding provides a simple yet efficient mechanism to generate a parity block by performing bitwise XOR across multiple data shards~\cite{luo2013efficient}. 
However, its resilience is limited to single-failure recovery, leaving the system vulnerable to correlated or multiple failures in large-scale clusters. 
row–diagonal parity (RDP)~\cite{corbett2004row} further introduces a lightweight, systematic XOR-based scheme that protects data using two parity shards, namely row parity and diagonal parity, enabling recovery from any two simultaneous shard failures via an efficient diagonal-walk reconstruction~\cite{goel2012raid}. 
In contrast, Reed–Solomon (RS) coding~\cite{rs} performs encoding and decoding over Galois fields using unique coefficients per parity symbol, providing higher fault tolerance but higher latency and storage costs.

\section{Motivation}
\label{sec:motivation}
While recent efforts have dramatically improved LLM serving performance, fault tolerance and reliability remain critical, yet often overlooked, aspects of system design.
In production, LLMs are deployed on multi-GPU and multi-node infrastructures where a single device failure can halt the entire serving process~\cite{burstgpt,dejavu}. 
A Microsoft study of 156 high-severity production incidents shows that around 60\% of failures occur at the inference engine level, including but not limited to KV cache errors, memory leaks, and crashes~\cite{reliabileai}. 
This new analysis echoes prior studies indicating GPU errors dominate the root causes for large-scale ML training applications~\cite{revisiting,robustlt}. 
As agent-based applications are increasingly embedded in our daily lives, processing million-token workloads has become a reality in production-level systems, where a single point of failure could lead to minutes or even hours of wasted time and resources~\cite{zhu2025sampleattention,cp}.

Unlike in LLM training, \textbf{the central challenge for reliable LLM serving is protecting the transient KV cache, which expands with sequence length and batch size.} 
This fragility is especially acute for long-context generation. 
A full restart of requests with millions of tokens to recompute the massive KV cache is expensive.
For instance, processing 1M context length for a 405B model can take 20 minutes on a single node with 8$\times$H100 GPUs~\cite{cp}.
Although traditional solutions like checkpointing are effective for distributed training, they are ill-suited for the unique demands of LLM serving due to two primary drawbacks:
\begin{itemize}
    \item \textbf{Prohibitive Latency Overhead.} 
    Checkpointing, while straightforward, introduces severe latency by stalling the prefill stage to write the KV cache state to host memory or disk. 
    This overhead is amplified in intra-node tensor parallelism, where the I/O operations cannot be effectively overlapped with computation. 
    As shown in Figure~\ref{fig:motivation2}, this can increase prefill latency by as much as 113\% for 70B models. 
    Such a bottleneck will be exacerbated as applications scale to process million-token contexts. 
    \textit{This necessitates a more efficient, lightweight method for protecting the KV cache with minimal performance impact}.
    \item \textbf{Excessive Memory Overhead.} 
    Prevalent checkpointing methods also suffer from high memory consumption. 
    While storing a full copy of the KV cache in host memory is simple, this brute-force replication can exhaust node memory, ironically degrading overall system reliability.
    In Figure~\ref{fig:motivation2}, we can see that the KV cache memory footprint can consume over 300\,GB for long-sequence inputs.
    Furthermore, in high-capacity serving scenarios, host memory often acts as a buffer for the KV cache of preempted requests~\cite{vLLM,fastserve}. 
    Adding a full checkpointed KV cache creates further resource contention.    
    \textit{Therefore, an efficient and flexible memory management and scheduling strategy for fault tolerance is crucial.}
    
\end{itemize}

\section{Methodology}

\begin{figure*}[!t]
     \centering
         \includegraphics[width=\linewidth]{ 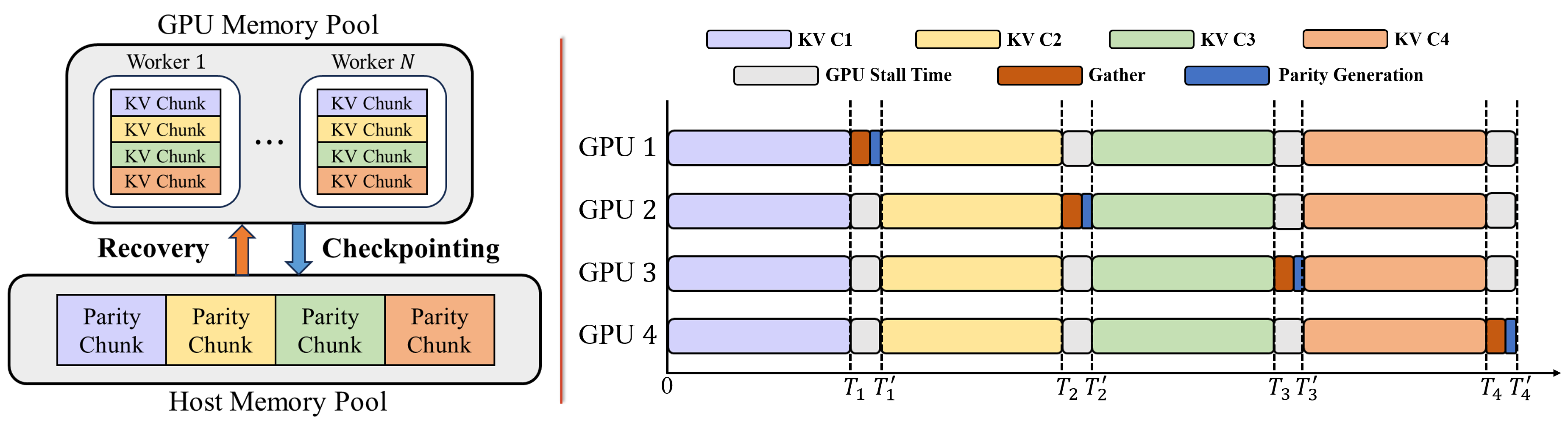}
         \vspace{-8mm}
        \caption{(a) Left: System overview of \ourmethod.
        (b) Right: Illustration of the chunk-level checkpointing and load-balancing strategy in terms of GPU execution timeline. 
        As KV cache chunks are generated ($T_1, T_2, T_3, T_4$), a different GPU is assigned in a round-robin fashion to gather and compute the corresponding parity chunk, thereby distributing the checkpointing overhead.
        The computation for subsequent chunks then resumes after checkpointing ($T_1^{'}, T_2^{'}, T_3^{'}, T_4^{'}$).
		}
	\label{fig:ghostserve-high-level}
\end{figure*}

In this work, we propose a new lightweight checkpointing system, \ourmethod, for distributed LLM serving based on the idea of erasure coding to address the fault-tolerance challenges outlined in Section~\ref{sec:motivation}.
The overview system architecture is shown in Figure~\ref{fig:ghostserve-high-level}~(a).
Compared to naive replication-based checkpointing, \ourmethod leverages erasure coding to generate parity shards for the KV cache at the granularity of a chunk (group of tokens).
The parity shards are stored in host memory and retrieved when one or multiple KV caches are lost during GPU failures.
To promote stability, \ourmethod also features a load-balancing strategy to assign parity generation workloads in a round-robin manner.

\subsection{Erasure Coding for LLMs}
\label{subsec:ec}

To mitigate checkpointing overheads, we employ erasure coding to protect the streaming KV cache during LLM serving. 
Unlike replication, which duplicates the entire KV cache, our approach stores only $K$ parity shards for $N$ data shards ($K \ll N$). 
If a failure occurs, lost data is reconstructed using the surviving data and parity shards from the remaining workers. 
This method substantially reduces storage costs.
As shown in Figure~\ref{fig:motivation2}, an 8:2 data-to-parity ratio reduces overhead by 75\% compared to full replication.

However, applying erasure codes directly to the KV cache is far from trivial. 
One challenge is the format incompatibility, as standard schemes operate on bit-wise parity, as opposed to the floating-point KV tensors. 
We address this by reinterpreting each FP16 tensor value as a fixed-width integer bit pattern based on the IEEE-754 standard~\cite{saikia2020efficient, mathis2022implementation}. 
This reinterpretation is fully reversible, ensuring lossless recovery of the original floating-point values, and it simplifies the design of efficient, low-latency GPU kernels.
Moreover, the unique structure of erasure codes like Reed-Solomon (RS) requires the tensors to be treated as coefficients of a polynomial over a finite Galois Field~\cite{rs}.
A direct PyTorch implementation would run into issues of excessive global memory access times, thus dominating the runtime and inducing significant latency overhead.

To this end, we design custom native CUDA kernels by fusing FP16-to-uint16 packing, parity computation, and unpacking into single GPU passes using 64-bit indexing, grid-stride loops, and warp-synchronous execution to maximize occupancy and avoid race conditions. 
Compared to conventional tensor-wise casting or CPU-assisted checksum computation, our approach eliminates format conversion overheads and intermediate memory transfers.
Consequently, our kernels deliver significantly lower latency than native PyTorch-based implementations, as shown in Figure~\ref{fig:microbenchmarks_combine}.
Furthermore, \ourmethod provides flexibility by implementing three standard erasure coding schemes, namely, XOR~\cite{aguilera2005using}, RDP~\cite{corbett2004row}, and Reed-Solomon~\cite{rs}, allowing users to select a protocol based on their fault-tolerance needs. 
In distributed tensor parallelism settings, \ourmethod encodes data across the $N$ data shards (devices) and can reconstruct data from up to $K$ simultaneous failures (parity shards). 
Detailed implementations can be found in Section~\ref{sec:implementation}.

\subsection{\ourmethod Scheduler}
\label{subsec:scheduler}
Our proposal to implement erasure coding provides flexible, low-latency fault tolerance for checkpointing and recovering the KV cache. 
While the concept is straightforward for single-GPU scenarios, its application in distributed settings is non-trivial, particularly for tensor parallelism. 
\textbf{The key problem is how to efficiently schedule the checkpointing process and manage the data/parity shards.}
A natural solution is to have each device perform erasure coding independently, storing necessary shards from peers in its local HBM memory.
However, such a strawman solution suffers from two main drawbacks.  

First, it has been shown that the performance of LLM serving highly depends on the availability of GPU memory~\cite{vLLM,fastserve,zheng2024sglang}.
Such a naive approach would induce significant GPU memory overhead, as each GPU would have to allocate additional space to store data or parity shards from other GPUs. 
This severely limits the serving capacity and may lead to a degradation in the quality of service (QoS).
Furthermore, coordinating the data and parity shards across GPUs involves complex communication protocols, which often require ring-based operations, such as \textit{all-to-all}. 
These additional communication overheads could lead to potential system bottlenecks.

\minisection{Chunk-level Checkpointing.}
To this end, we propose to perform checkpointing at the granularity of a chunk, defined as a group of tokens.
Instead of conducting erasure coding on each device individually, we collect all the KV cache and view each KV cache as an individual data shard, and generate corresponding parity shards.
An illustration of the above process is shown in Figure~\ref{fig:ghostserve-high-level}. 
After each KV cache chunk is generated ($T_1, T_2, T_3, T_4$), a GPU is specified to gather the KV cache chunk and conduct the parity generation workloads.
The GPUs are then synchronized after parity generation to resume the inference process for the next chunk. ($T_1^{'}, T_2^{'}, T_3^{'}, T_4^{'}$).
Thanks to fast intra-node NVLink and our fast erasure coding kernel, the idle time in between each chunk accounts for less than 5\%.
These parity shards are subsequently offloaded to host memory, thus eliminating GPU memory overhead.
Unlike the strawman, which requires large, persistent HBM allocations on each GPU to store peer shards, our approach only requires a small, fixed-size temporary buffer for the gather operation. 
This process can be seamlessly integrated into chunked prefill and, by operating at the chunk level, also minimizes its impact on decode throughput.

\begin{algorithm}[!t]
\caption{GhostServe Checkpointing Workflow.}
\label{algo:checkpointing}
\begin{algorithmic}[1]
\Statex \textbf{Initialization:} 
Initialize erasure coding compute unit $\textit{EC}$, KV cache $KV$,
chunk size $m$, 
input length $s$,
number of prefill chunks $c=\lceil s/m \rceil$,
the number of GPUs $N$, number of parity shards $K$,
GPU assignment index $k$.
\State $i, k=0, 0$ 
\For{all $i < c$} 
\For{all $j < N$} 
\State \textcolor{gray}{\# Generate KV cache chunk}
\State $KV_{i}^{j} = \text{GPU$^{j}$.Process}(m, i)$
\EndFor
\State \textcolor{gray}{\# Gather KV cache to $k$-th GPU}
\State $KV_{i} = \text{torch.dist.gather}(KV_{i}^{1:N}, \text{GPU$^{k}$})$
\State \textcolor{gray}{\# Perform erasure coding}
\State $P_{i} = \text{EC.encode}(KV_{i})$
\State \textcolor{gray}{\# Transfer parity shards to host memory}
\State $P_{i} \xrightarrow[\text{Async}]{\text{PCIe}} \text{DRAM}$
\If{$k==N$}
\State \textcolor{gray}{\# Reset assignment index} 
\State $k=0$ 
\Else
\State \textcolor{gray}{\# Move to the next GPU for encoding}
\State {$k=k+1$}
\EndIf
\EndFor
\end{algorithmic}
\end{algorithm}

\minisection{Recovery with Partial Recomputation.}
To further speed up the recovery process, \ourmethod employs a hybrid strategy, where partial preceding KV cache is recomputed.
Upon failure, the system initiates GPU-side recomputation only for the initial portion of the KV cache, while the remaining segments are recovered using the available parity and surviving data shards. 
This hybrid mechanism eliminates the need for GPUs to remain idle until all parity shards are transferred from CPU to GPU memory. 
Instead, it overlaps recomputation with I/O transfers, effectively hiding latency and ensuring continuous GPU utilization during the recovery process.

\minisection{Load Balancing.}
A static assignment, where one GPU is repeatedly tasked with parity computations, would create potential straggler effects, inducing severe workload imbalance and performance bottleneck~\cite{SI,SI2}. 
To prevent `hotspots' and reduce the risks of single-device thermal wear-out, \ourmethod features a load balancing strategy that distributes the parity computation load, as shown in Figure~\ref{fig:ghostserve-high-level}~(b).
The duty is rotated across devices in a round-robin manner (e.g., GPU 0 computes parity for Chunk 0, GPU 1 for Chunk 1, and so on). 
This strategy ensures that the computational and communication overhead of erasure coding is shared equally, promoting balanced GPU utilization. 
Redundancy is generated on-the-fly without intermediate memory staging, enabling high GPU occupancy and sustained NVLink bandwidth for efficient, fault-tolerant LLM serving.

\minisection{Overall Workflow.}
Combining the above-mentioned techniques, we now describe the workflow of \ourmethod scheduler in detail, including the checkpointing and recovery process.

1) \textbf{Redundancy-based Checkpointing.} We present the detailed procedure of our checkpointing process in Algorithm~\ref{algo:checkpointing}.
For readability, we describe the checkpointing only for the prefill, since the same process can be applied for decoding, where the parity is updated once the KV cache for a chunk of tokens is generated.
GhostServe employs dynamic chunking, where a request with an input length $s$ is partitioned into $\lceil s/m \rceil$ chunks, and the final partial chunk is handled using thread-masking and bounds-checking within our CUDA kernels, where the parity is only computed for active KV caches.
In lines 3-6, for each input chunk, all $N$ GPUs work in parallel to compute their respective portions of the KV cache.
In Lines 8-13, once a full KV cache chunk has been generated across all GPUs, we first collect the distributed pieces of the chunk, denoted as $KV_i^j$ on a single, designated GPU, indexed by $k$, using \textit{torch.dist.gather}.
Next, \text{GPU$^{k}$} uses an encoding function $\text{EC.encode($\cdot$)}$ to compute a set of redundant parity shards for the entire chunk.
The newly created parity shards are immediately moved to the main system's host memory.
Lastly, to ensure balanced workloads, the GPU assignment index is updated at every chunk until all input tokens are processed.

2) \textbf{Hybrid Recovery.} 
We also present the design of our recovery procedure in Algorithm~\ref{algo:recovery}. 
We use failure during prefill and a single-GPU failure as examples for simplicity.
Failure scenarios for both decoding and multi-GPU can be readily generalized, as the recovery function is applied to the failed GPUs separately.
Compared to prior checkpointing methods, we adopt a hybrid recovery mechanism, integrating recomputation to further reduce stall time.
\begin{algorithm}[!t]
\caption{GhostServe Recovery Workflow.}
\label{algo:recovery}
\begin{algorithmic}[1]
\Statex \textbf{Initialization:} 
Initialize erasure coding compute unit $\textit{EC}$, KV cache $KV$, 
chunk size $m$, 
input length $s$,
number of prefill chunks $c=\lceil s/m \rceil$,
the number of GPUs $N$, number of parity shards $K$.
Assuming $k$-th GPU failed after $n$-th chunk.
\State ...
\State \text{Failure.detect()}
\State \textcolor{gray}{\# Calculate the recompute units}
\State $r=\text{get\_recompute\_units}(s,m,N,K)$
\If{$r>=n$}
\State \textcolor{gray}{\# Perform recomputation} 
\For{all $i < n$}
\State $KV_{i}^{k} = \text{GPU$^{j}$.Process}(s, i)$
\EndFor
\Else
\State \textcolor{gray}{\# Perform recomputation + reconstruction} 
\For{all $i < r$}
\State $KV_{i}^{k} = \text{GPU$^{j}$.Process}(s, i)$
\EndFor
\State \textcolor{gray}{\# Gather remaining KV cache to $k$-th GPU}
\State $KV_{1:n-r} = \text{torch.dist.gather}(KV_{1:n-r}^{1:N}, \text{GPU$^{k}$})$
\State $P_{1:n-r} \xrightarrow[\text{Async}]{\text{PCIe}} \text{GPU$^{k}$}$
\State \textcolor{gray}{\# Perform reconstruction}
\State \text{With CUDA.Streams():}
\State \quad $KV_{1:n-r} = \text{EC.reconstruct}(KV_{1:n-r}, P_{1:n-r})$
\EndIf
\State \text{Serving.Resume()}
\State ...
\end{algorithmic}
\end{algorithm}
During interruption, the process begins when a failure is detected on a specific GPU, e.g., $k$-th GPU after a certain number of chunks ($n$-th chunk) have been processed. 
In lines 3-4, we first compute the optimal number of chunks ($r$) that should be recomputed from scratch.
For short sequences, full recomputation is conducted for recovery, which avoids the communication overhead of gathering data for erasure coding reconstruction (see lines 5-9).
In the case of long-context scenarios, erasure coding is further applied.
Apart from recomputing $r$ KV cache chunks, the scheduler gathers $(n-r)$ chunks of data shards from surviving GPUs and transfers the parity shards from CPU memory to the failed GPU in an asynchronous fashion.
Next, the reconstruction function $\text{EC.reconstruct($\cdot$)}$ for erasure coding is applied for each chunk individually with CUDA streams (lines 18-20).
Once the failed GPU state has been fully restored through either recomputation or reconstruction, the inference job resumes from the interruption point.

We note that \ourmethod is primarily designed for intra-node serving, thus focusing on individual device memory software faults, such as silent data corruptions~\cite{sdc,sdc2}, memory errors~\cite{stg}, kernel faults~\cite{ft-fsdp}, and resource leaks~\cite{cuda-leaks}, which do not usually require a hard restart of the entire node.
In this work, we only consider scenarios where the failed GPUs can be restarted to rejoin the surviving GPUs to resume inference services.
In practice, the recovery process also includes re-establishing NCCL connections, GPU warmups, and CUDA graph capture, which should occur before KV cache recovery.
Failure scenarios for both decoding and multi-GPU can be readily generalized, as the recovery function is applied to the failed GPUs separately.
\ourmethod can tolerate up to $K$ number of simultaneous GPU failures, same as the number of parity shards, without resorting to pure recomputation.

\section{Implementation}
\label{sec:implementation}
\ourmethod is an end-to-end system implemented with 4K lines of Python and 1.5K lines of C++/CUDA. 
We use SGLang version 0.5.1~\cite{zheng2024sglang} as our backend and implement our method as a plug-in module. 
In practice, \ourmethod can be integrated into existing serving engines such as vLLM \cite{vLLM} and HuggingFace-TGI~\cite{huggingface} with minimal modifications, making it portable and easy to adopt. 
For the attention backend, we adopt FlashInfer version 0.3.1~\cite{flashinfer}, built on top of PyTorch 2.8 with CUDA 12.6.
We use NCCL version 2.21.5 for GPU communication within the node.
Our code is available here\footnote{\small{\url{https://github.com/project-ghostserve/26mlsys-AE-GhostServe}}}.

\begin{figure*}[!t]
         \includegraphics[width=1\linewidth]{ 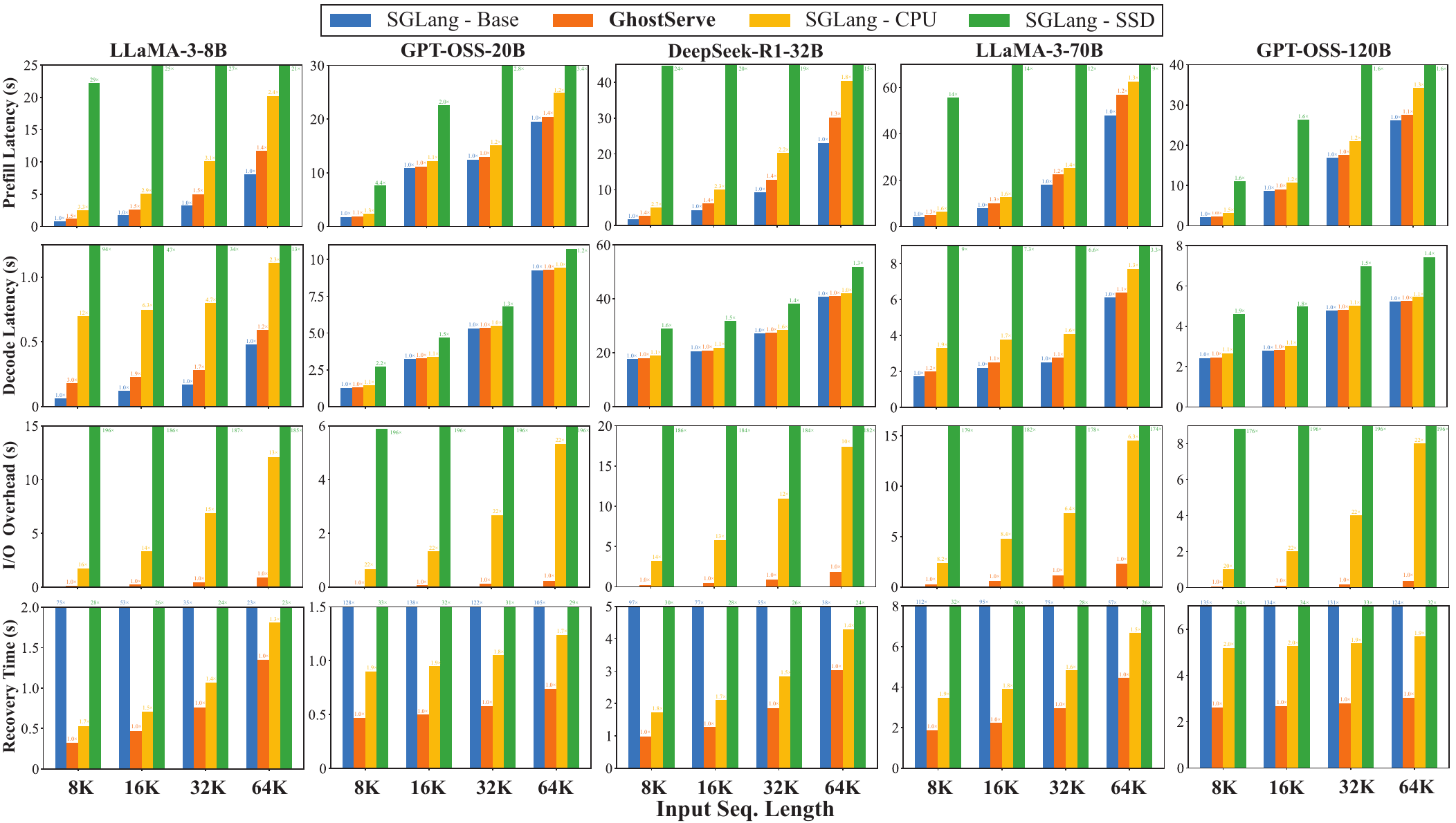}
         \vspace{-8mm}
        \caption{Performance comparison of different fault-tolerant methods. 
        The latency results are measured with a batch size of 16, chunk size of 2K, and an output length of 4K, using varying input lengths that range from 2K to 64K.
        I/O overhead represents the total I/O latency incurred during the checkpointing process. 
        Recovery latency denotes the time required to restore 50\% of the chunks to resume inference.
        All models are running with an 8:2 parity ratio for \ourmethod.
        }
     \label{fig:main_results}
\end{figure*}

\minisection{Kernel Optimization.} 
Since \ourmethod introduces additional compute overhead, achieving high performance requires carefully optimized GPU kernels. 
We implement three key optimization techniques:
(1) \textit{Kernel fusion.}
We reduce the kernel launch overhead by fusing multiple operations into a single GPU kernel. 
Instead of separately invoking the conversion, encoding, and reconstruction operations, we perform kernel fusion for both the checkpointing and recovery procedures in a single pass.
(2) \textit{CUDA Graph.} 
As the erasure coding is conducted at the chunk level with a fixed size, we further leverage the CUDA graph to minimize kernel launch overhead, where graphs of encoding and reconstruction operations are captured and replayed.
(3) \textit{CUDA Streams.} To accelerate recovery when multiple chunks of the KV cache must be reconstructed concurrently, \ourmethod employs a multi-stream CUDA execution strategy. 
Each chunk is assigned to a dedicated CUDA stream, enabling concurrent parity decoding fully utilizing the available GPU SMs.

\minisection{Continuous Batching.}
While \ourmethod is primarily designed for long-input batched inference, we also extend GhostServe for online serving by creating a CUDA stream pool to manage each request. 
During the forward checkpointing process, each incoming request is assigned a CUDA context, which generates and updates the parity individually. In the event of KV cache loss, multiple streams are launched to recover the KV cache for each request, allowing the GPU to overlap kernel executions.

\section{Evaluation} 
\subsection{Experimental Setup}
\minisection{Hardware.} 
We conduct our experiments on a private cluster, configured with 8 NVIDIA H200 GPUs, connected over NVLink Gen 4, and two 48-core Intel processors, with 1024 GB DDR5 memory, connected through PCIe Gen 4 with a maximum bidirectional bandwidth of 32 GB/s.

\minisection{Models.} 
We evaluate \ourmethod on a diverse range of LLMs to demonstrate its generalization across model types and scales. 
Specifically, we use LLaMA-3-8/70B~\cite{llama}, GPT-OSS-20/120B~\cite{openai2025gptoss}, and DeepSeek-R1-32B~\cite{guo2025deepseek}. 
All models are executed in half precision (FP16). 

\minisection{Workloads and Metrics.}
We synthesize traces using the Medha generator~\cite{medha} to generate a mix of long-input-short-output and short-input-long-output requests, with various batch sizes (4$\sim$64) and sequence lengths (4K$\sim$1M). 
For request timestamps, we model request arrival times according to a Poisson distribution.
For efficiency metrics, we use prefill, decode, and recovery latency for batched inference, and P50 and P99 latency for online serving.
In terms of cost-benefit analysis, we use 
(1) \textit{Effective-Inference-Time-Ratio} (EITR): Defined as the division of actual inference time and total runtime. 
This helps quantify the overhead/costs for checkpointing.
(2) \textit{Mean-Time-To-Recover} (MTTR): Defined as the average recovery time for entire request traces. This helps quantify the benefits of checkpointing.

\begin{figure*}[!t]
         \includegraphics[width=1\linewidth]{ 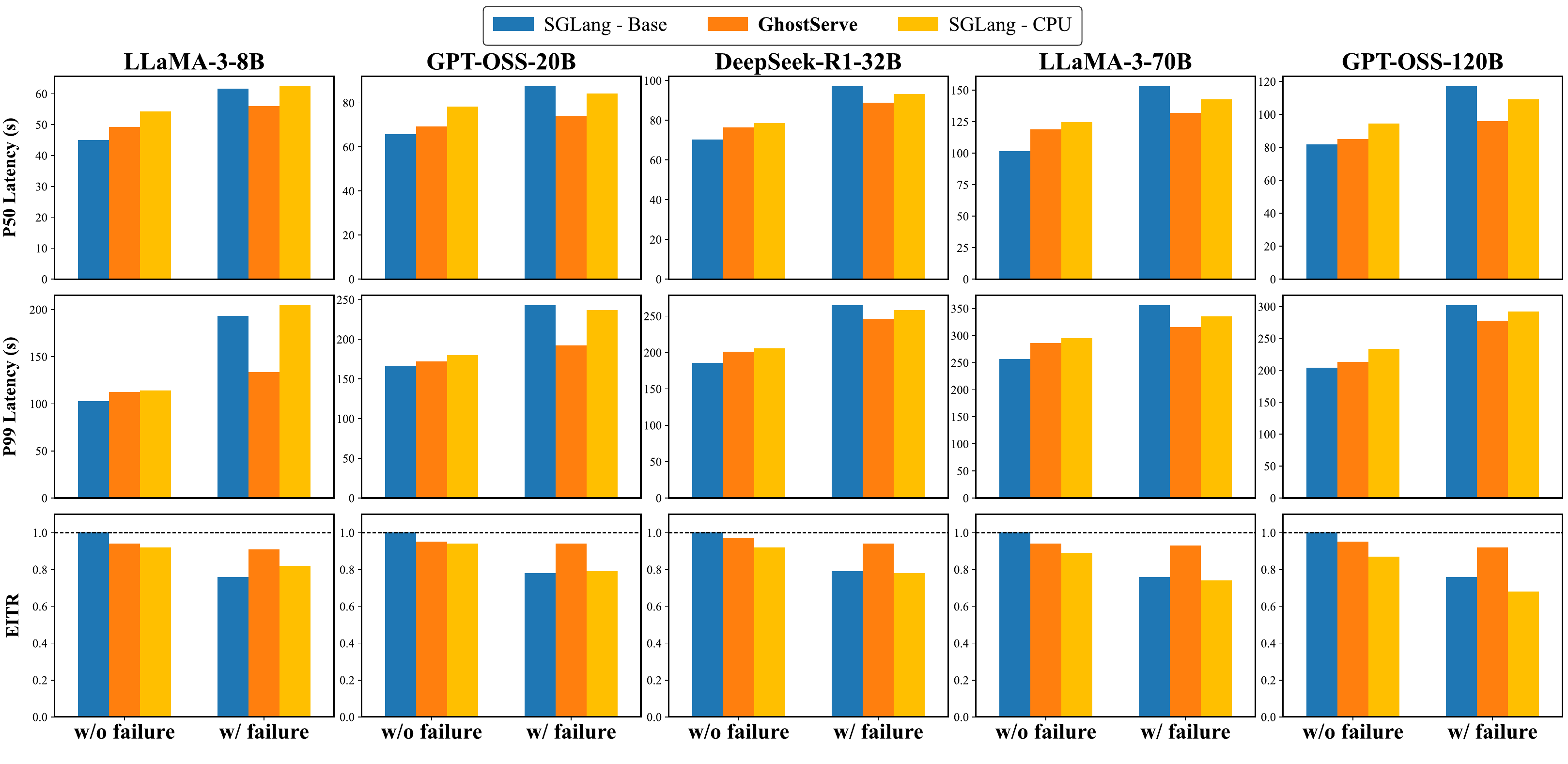}
         \vspace{-9mm}
        \caption{
        Performance comparison of different fault-tolerant methods in online serving. 
        Here, we measure P50/99 latency and effective-inference-time-ratio (EITR) under both failure-free and failure-induced environments.
        Faults are injected at random steps with a failure rate 15\%.
        } 
     \label{fig:online_results}
\end{figure*}

\minisection{Failure Simulation.}
In terms of fault injection, we first build a failure model using statistics from prior work~\cite{dejavu,reliabileai,burstgpt,robustlt}.
We mimic device faults by completely flushing the memory buffers of particular workers and hanging the other workers until all the data is recovered.
Following prior works~\cite{reliabileai,burstgpt}, we vary the overall failure rate from 5\% to 15\%, and inject failure at random points throughout the request's execution runtime.

\minisection{Baselines.} We compare \ourmethod with three previous baseline methods:
\begin{itemize}
    \item \textbf{SGLang - Base}: The majority of LLM serving engines lack a sufficient recovery mechanism for the KV cache, and employ a simple recomputation technique when an interruption occurs~\cite{fastertransformer,vLLM,zheng2024sglang}.
    We refer to this baseline as SGLang - Base.
    \item 
    \textbf{SGLang - SSD}: 
    Existing training-based methods provide a mechanism to checkpoint the model weights to persistent storage, such as NVMe SSD. 
    Here, we reproduce a similar asynchronous baseline based upon the state-of-the-art PCCheck~\cite{pccheck}, and refer to SGLang - SSD.
    \item
    \textbf{SGLang - CPU}: 
    D{\'e}j{\`a}Vu is the state-of-the-art LLM fault-tolerant solution for multi-GPU serving, which utilizes a state replication method to store KV cache in CPU memory~\cite{dejavu}.
    Here, we implement D{\'e}j{\`a}Vu for SGLang in an asynchronous manner for intra-node tensor parallelism, where we refer to SGLang - CPU.
\end{itemize}

\subsection{End-to-End Results}
\begin{figure*}[!t]
     \centering
          \includegraphics[width=1\linewidth]{ 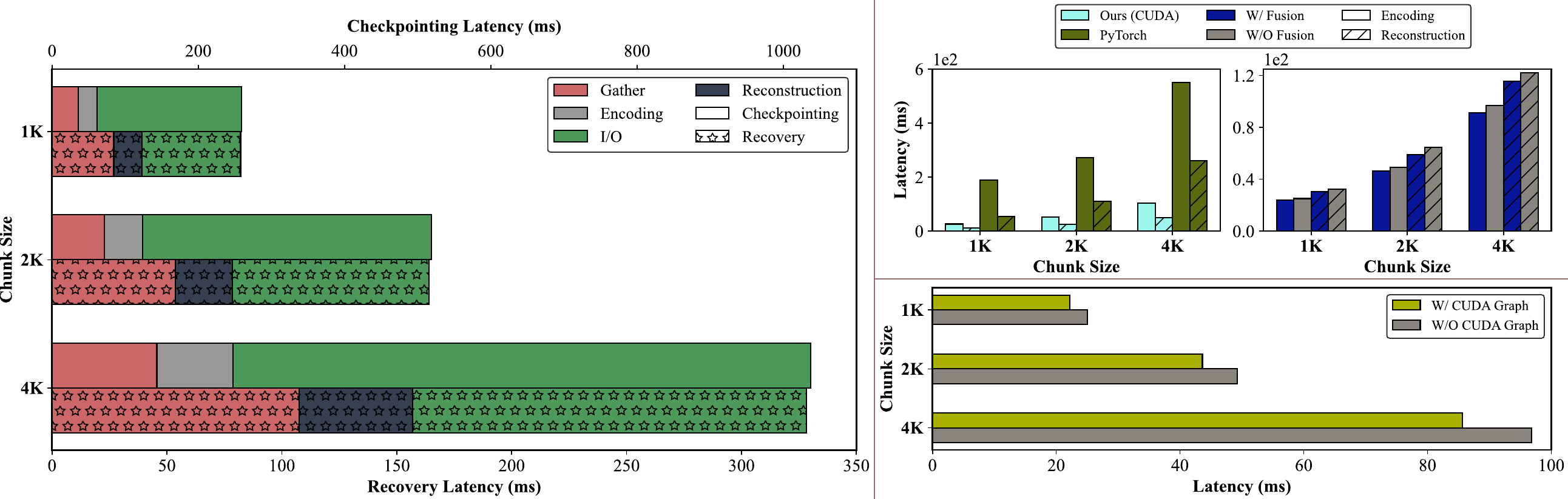}
         \vspace{-8mm}
        \caption{ Kernel Microbenchmark.
        All experiments are conducted using LLaMA-3-70B with a batch size of 16.
         (a) Left: Performance breakdown for erasure coding kernel during checkpointing and recovery for different chunk sizes.
         (b) Right: Impact of implementation method (PyTorch vs CUDA), kernel fusion, and CUDA graph on the erasure coding performance for different chunk sizes.
         }
     \label{fig:microbenchmarks_combine}
\end{figure*}

\minisection{Batched Inference.}
Figure \ref{fig:main_results} demonstrates the performance comparison across different models and scales with varying input sequence lengths for different methods.
We observe that \ourmethod shows significant improvements against prior checkpointing methods.
Three key insights emerge from these results.
First, \ourmethod consistently yields lower latency overhead during prefill.
It delivers an average 2.7$\times$ speedup over CPU-based replication, despite its complex procedure.
Second, it also has a minimal impact on the decode latency, inducing less than a 10\% overhead. 
\ourmethod delivers a dramatic 47$\times$ speedup over SSD-based replication.
This efficiency stems from our efficient erasure coding kernel operation and reduced checkpointing memory overhead.
It can be seen that \ourmethod achieves a 13$\times$ reduction in I/O overhead relative to CPU checkpointing and a 132$\times$ reduction compared to SSD checkpointing. 
Third, \ourmethod consistently achieves the lowest recovery latency, even on large models such as GPT-OSS-120B and LLaMA-70B, while competing approaches incur orders of magnitude higher costs. 
For instance, on LLaMA-3-70B with 64K input tokens, \ourmethod recovers in under 5 seconds, compared to close to 2 minutes for the SSD-based method. 
These results confirm that GPU-centric parity generation substantially minimizes checkpointing latency and memory overhead, making \ourmethod far more practical for large-scale fault-tolerant LLM serving.
\begin{figure}[!t]
     \centering
         \includegraphics[width=\linewidth]
         { 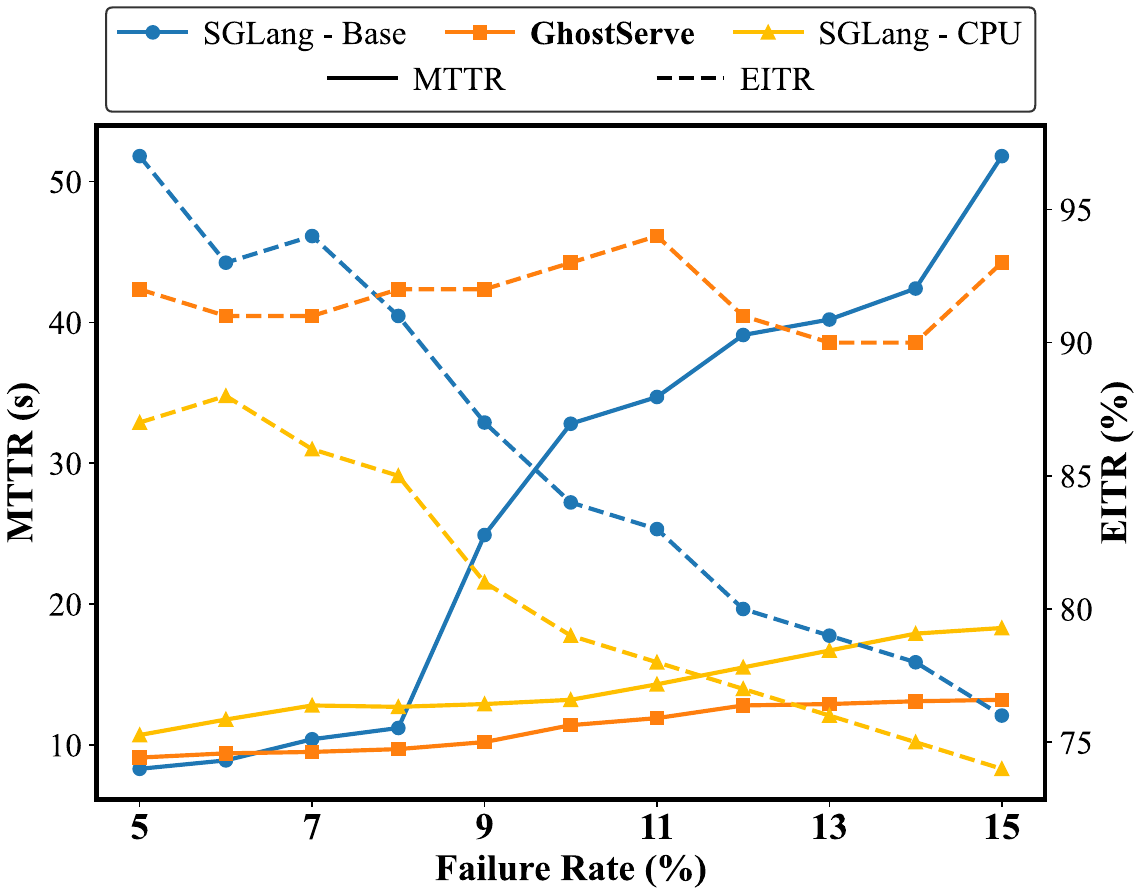}
         \vspace{-8mm}
        \caption{
        Cost-benefit analysis for serving LLaMA-3-70B model over the entire serving traces.
        Here, we compare the EITR and MTTR for different methods under varying failure rates (5\%$\sim$15\%).
        }
	\label{fig:reliability}
\end{figure}

\minisection{Online Serving.}
Figure~\ref{fig:online_results} further shows the efficacy of our method in continuous serving scenarios.
Specifically, we compare how each method affects the normal inference process.
Here, we summarize three key observations.
\underline{First}, in failure-free settings, \ourmethod provides much lower checkpointing overheads in terms of mean and tail latency, with up to 11\% reduction.
\underline{Second}, \ourmethod significantly mitigates the effect of interruptions or failures during serving.
For instance, in the LLaMA-70B model, thanks to the fast recovery mechanism of erasure coding, \ourmethod achieves 1.2$\times$ and 1.1$\times$ speedups in P50 and P99 latency against naive recomputation, respectively.
\underline{Third}, \ourmethod achieves consistently high EITR ($>90\%$) compared to baseline methods. 
In particular, \ourmethod improves upon replication by an average of 23\% under failures.
The benefits of CPU-checkpointing diminish as the model size scales up, where for larger models, such as 70B and 120B, it actually underperforms the recomputation baseline, due to its high I/O overheads, further demonstrating the cost-effectiveness and practicality of \ourmethod in real-world serving.

\subsection{Performance Analysis}
\begin{figure*}[!t]
     \centering
         \includegraphics[width=1\linewidth]{ 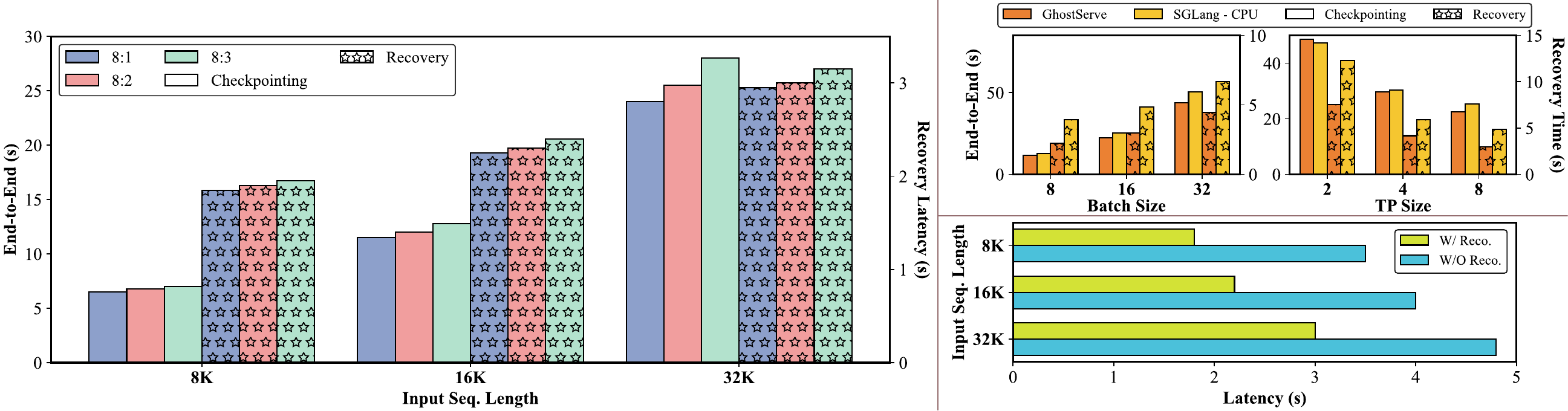}
         \vspace{-8mm}
        \caption{Sensitivity Studies. 
        All experiments are conducted using LLaMA-3-70B with a chunk size of 2K.
        Recovery latency is the time required to restore 50\% of the KV cache.
        (a) Left:
        Performance comparison of \ourmethod under different parity ratios. 
        (b) Top Right:
        Performance comparison of different fault-recovery methods with varying batch sizes and TP sizes. 
        (c) Bottom Right:
        Ablation studies on the impact of recomputation on the recovery latency.
        } 
     \label{fig:sensitivity}
\end{figure*}

\minisection{Cost-Benefit Analysis.}
Figure~\ref{fig:reliability} demonstrates the detailed reliability analysis of \ourmethod on serving LLaMA-3-70B.
Here, we summarize two observations.
First, \ourmethod can maintain relatively high EITR under varying failure rates, showcasing its robustness in different scenarios.
In contrast, both baseline methods suffer from performance degradation as the failure rate rises.
Second, \ourmethod provides much lower mean-time-to-recover (MTTR) than replication across different settings.
This is thanks to the hybrid approach of \ourmethod, which combines both recomputation and erasure coding, operating in parallel to speed up the KV cache recovery.

\minisection{Kernel Microbenchmark.}
Figure~\ref{fig:microbenchmarks_combine} presents the kernel-level latency breakdown of \ourmethod, revealing three key observations.
First, the overhead from collection and erasure coding is modest, remaining significantly lower than the dominant GPU-to-CPU I/O transfer time. 
Second, parity generation and reconstruction complete faster than the NCCL operations used for KV cache collection, and the reconstruction process takes less time than the NCCL operations to collect the KV cache.
One reason for this is the inherent nature of many-to-one \textit{torch.dist.gather} operation, requiring all GPUs to be synchronized.
Third, our custom kernel delivers significantly lower latency than a native PyTorch-based implementation.
Kernel fusion and CUDA graphs further accelerate the erasure coding by up to 1.05$\times$ and 1.13$\times$ on encoding and reconstruction, respectively, across different chunk size configurations.

\minisection{Sensitivity Studies.} 
Figure~\ref{fig:sensitivity} presents the results of \ourmethod for further performance analysis for different parity ratios, batch sizes, and the impact of recomputation in our hybrid recovery procedure.
Here, we summarize four key insights.
First, \ourmethod remains robust under different fault-tolerance requirements, where increasing the parity ratios adds marginal overheads to both checkpointing and recovery latency, indicating the scalability of our method.
Second, similar to sequence length scaling, \ourmethod outperforms prior methods consistently in terms of checkpointing overheads and recovery latency, and scales well with batch sizes.
Third, \ourmethod consistently outperforms CPU-checkpointing in high TP settings (TP$>$2).
For TP$=$2, the benefit of erasure coding vanishes due to its extra overhead in encoding and no reduction in I/O transfer latency.
Last, we observe that our hybrid recovery mechanism of combining recomputation and reconstruction results in significant recovery time improvement, achieving up to 42.9\%.
This stems from the reduced amount of host-device data transferred and computation load for reconstruction, thereby lowering recovery latency.

\begin{figure}[!t]
     \centering
         \includegraphics[width=.95\linewidth]{ 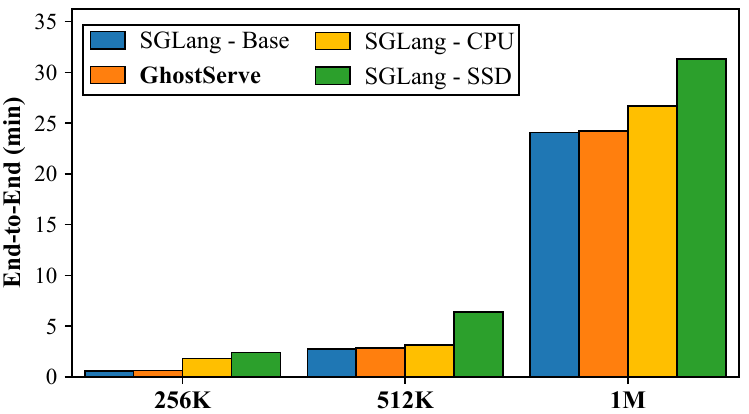}
         \vspace{-4mm}
        \caption{Performance comparison of different methods when scaling to million-tokens. 
        Results are reported using LLaMA-3-70B, with a batch size of 1, chunk size of 2K, and output length of 4K.
        }	
	\label{fig:1M}
\end{figure}

\minisection{Scaling to Million Tokens.}
We further conduct experiments with extremely long sequences, up to 1M tokens, as shown in Figure~\ref{fig:1M}. 
We make two key observations.
First, \ourmethod induces less than 6\% overhead upon the baseline method, highlighting the scalability of our method in real-world, agent-based workloads.
Moreover, compared to D{\'e}j{\`a}Vu, our method significantly reduces the checkpointing overhead.
For instance, in 1M prefill, the overhead drops from 2.6 mins to only 9 seconds.
This demonstrates that our solution is far more practical for production-level serving systems when dealing with long-sequence inputs.

\section{Related Work}
\minisection{LLM Serving Systems.}
A large body of work focuses on improving LLM serving for high throughput and low latency through efficient KV cache management~\cite{vLLM,alisa,alise}, and kernel optimization~\cite{flashinfer,flashattention2}. 
Notably, vLLM introduces \textit{paged} key-value (KV) cache to improve memory efficiency and reduce fragmentation~\cite{vLLM}. 
SARATHI proposes chunked-prefill and stall-free batching to overlap prefill and decode stages to improve GPU utilization.~\cite{agrawal2023sarathi}.
Kernel libraries, such as FlashAttention~\cite{flashattention2} and FlashInfer~\cite{flashinfer}, aim to improve the execution speed at the operator level.
Nonetheless, these systems do not consider the aspect of fault tolerance and resort to naive recomputation for KV cache recovery.
Furthermore, methods like D{\'e}j{\`a}Vu induce too much overhead for these high-performance systems in tensor parallelism settings.
To bridge this gap, \ourmethod leverages erasure coding to enable lightweight checkpointing for distributed LLM serving.

\minisection{Fault-Tolerance for LLMs.}
Providing fault-tolerance support during LLM training has become increasingly critical as model scales and training durations continue to grow~\cite{mohan2021checkfreq,robustlt,moetion,torchft}. 
CheckFreq is among the first to enable automatic checkpointing for deep neural networks (DNNs) at the iteration level~\cite{mohan2021checkfreq}.
PCCheck further leverages concurrent checkpointing to minimize overhead and recovery time~\cite{pccheck}.
For large-scale LLM training, ByteCheckpoint~\cite{bytecheckpoint} develops a unified library to support checkpointing at the scale of tens of thousands of GPUs.
MoEtion proposes a distributed, sparse in-memory checkpointing for mixture-of-expert (MoE) model training~\cite{moetion}.
TorchTitan offers a unified library with step-level fault tolerance for training LLMs at scale~\cite{torchft}.
Despite these efforts, traditional replication-based checkpointing is not optimal for LLM serving. 

\minisection{Fault-tolerance in Distributed Systems.}
Fault tolerance has long been a fundamental concern in distributed systems, where reliability is critical in the presence of hardware faults, network disruptions, or node crashes. 
Classic redundancy-based techniques, such as replication and checkpointing, provide strong protection but often incur prohibitive storage and performance overheads at scale. 
Erasure coding, particularly Reed–Solomon (RS)~\cite{guruswami2016repairing} and its variants, has emerged as a more storage-efficient solution and has been widely adopted in large-scale storage infrastructures. 
Beyond storage, distributed computing frameworks such as MapReduce~\cite{dean2008mapreduce} and MPI-based high-performance systems integrate checkpoint/restart and log-based recovery to mitigate task or node failures. 
In this work, we aim to leverage the classical idea of erasure coding to promote reliability for LLM serving.

\section{Discussion and Limitation}

\minisection{Cross-node Scalability.}
\ourmethod can be further extended to cross-node environments through hierarchical fault-tolerance coordination and bandwidth-aware parity placement. 
For inter-node redundancy, \ourmethod can designate one or more parity coordinators that aggregate parity blocks across nodes using NCCL over InfiniBand or NIC. 
The challenge of providing node-level reliability lies in the bandwidth disparity between hierarchical interconnects. 
High-speed intra-node links (e.g., NVLink) are constrained by slower inter-node fabrics.
Different parallelism strategies also affect both the placement and communication patterns of the distributed KV cache.
Future work must address this by developing a topology-aware communication strategy that intelligently schedules data transfers. 
Furthermore, remote storage disks must come into play in multi-node environments, due to their durability to sustain node-level failures.
Potential solutions should include efficient scheduling algorithms and implementations that dynamically select the most efficient communication paths for checkpointing and recovery.

\minisection{Full-stack Fault Tolerance.} 
While \ourmethod provides lightweight and efficient redundancy for KV cache protection, it does not achieve full-stack fault tolerance across the GPU runtime. 
In the event of a GPU failure, \ourmethod successfully restores the missing KV cache using erasure-based recovery, but the underlying parallelism topology remains static. 
This means that failed GPUs cannot be dynamically excluded or replaced during ongoing inference, as the NCCL communication graph and tensor partitioning are initialized at launch time~\cite{shoeybi2019megatron}. 
Without runtime reorganization of TP groups or model weight redistribution, inference must pause until the failed GPU recovers or the system creates a new topology. 
Consequently, \ourmethod primarily targets data-level GPU memory `soft' errors that do not induce hard system failure.
For hardware faults that cannot be recovered with system reboots, GPU resource overprovisioning~\cite{gpuo1,revisiting} is often a must to ensure minimal impact and real-time live service migration.
Integrating \ourmethod with full-stack frameworks like TorchFT~\cite{torchft} would enable end-to-end resilience, bridging the gap between memory-level protection and system-level fault recovery.

\minisection{Real-world Serving.}
While \ourmethod has shown promising results in serving long-input prefill-heavy workloads, its application to decode-heavy reasoning workloads requires further investigation.
A key challenge lies in the unpredictability of decode lengths, thus complicating the encoding protocol.
Furthermore, as prefill-decode (PD) disaggregated architectures have become mainstream for real-world serving~\cite{Qin2025MooncakeTM,distserve}, how to provide fault tolerance for both prefill and decode workers remains an open question.

\section{Conclusion}
This work identifies the fault-tolerance issues for LLM serving and proposes erasure coding to achieve low-latency checkpointing and recovery in the presence of system faults.
Built upon chunk-level checkpointing and load balancing techniques, we design and implement a system solution, termed \ourmethod, and evaluate it across different workloads and model scales.
Extensive experiments show that \ourmethod consistently outperforms existing methods, achieving up to 2.1$\times$ and 2.7$\times$ latency improvement in checkpointing and recovery, respectively, and up to 1.2$\times$ median response latency speedup.

\section*{Acknowledgments}
We want to thank anonymous MLSys reviewers for their constructive feedback and our Shepherd, Mark Zhao, for guiding our revision process.
This work was sponsored in part by the Lambda Research Grant and the U.S. National Science Foundation (NSF) under Grants 1907765, 2400014, and 2426368.
This work also used Delta at UIUC NCSA through allocation CIS250367 and 250473 from the Advanced Cyberinfrastructure Coordination Ecosystem: Services \& Support (ACCESS) program, which is supported by U.S. NSF grants 2138259, 2138286, 2138307, 2137603, and 2138296.
Authors Youpeng Zhao and Jun Wang are inventors on a pending patent application (App. No. 18/123,456)~\cite{patent}.

\bibliography{ref}
\bibliographystyle{mlsys2025}





\end{document}